# Machine Learning based Data Driven Diagnostic and Prognostic Approach for Laser Reliability Enhancement

Khouloud Abdelli[1,2], Helmut Grießer[1], and Stephan Pachnicke[2]

[1]ADVA Optical Networking SE, Fraunhoferstr. 9a, 82152 Munich/Martinsried, Germany
[2] Christian-Albrechts-Universität zu Kiel, Kaiserstr. 2, 24143 Kiel, Germany

E-mail: KAbdelli@adva.com

**ABSTRACT**

In this paper, a data-driven diagnostic and prognostic approach based on machine learning is proposed to detect laser failure modes and to predict the remaining useful life (RUL) of a laser during its operation. We present an architecture of the proposed cognitive predictive maintenance framework and demonstrate its effectiveness using synthetic data.

1. **Introduction**

Due to their versatility, lasers are employed in various fields including optical communications, military and medical applications. As reliability is crucial in many of those fields, significant attention has been devoted to designing and producing highly reliable laser devices. However, meeting this target while satisfying other stringent constraints like cost and size, can lead to increased maintenance costs or lower production yield. Hence, it is desirable to develop a robust prognostics and health management (PHM) system to reduce maintenance costs and outage risks, while ensuring laser reliability and availability.

PHM is an engineering process of monitoring the health of the device, detecting anomalies, diagnosing faults and predicting the remaining useful life (RUL). It is widely recognized as an efficient and practical approach to cope with the different reliability engineering challenges such as unplanned system downtime or unexpected equipment failure [1]. PHM methods can be broadly categorized into model-based and data-driven approaches. The first approach utilizes the knowledge about the system, the operating conditions and the life cycle along with the deep understanding of the physical phenomena including the degradation process to model the failure/degradation behavior of the system using explicit mathematical equations [2]. Although this approach is precise, developing a highly accurate model using this methodology is very costly, time consuming and computationally intensive. The second approach extracts useful insights from historical data (training data) to learn the degradation trend and to predict the future health state without requiring any specific knowledge or using any physical model. An easy and quick implementation and deployment, low cost and the fact that no knowledge is required about the system make this approach the preferable PHM methodology [2]. Nevertheless, a drawback of this method is the reliance on the quality and the quantity of the data available to train the data-driven model [3]. Insufficient or imbalanced data may affect the performance of the model in terms of precision, accuracy and generalization capabilities to new conditions.

In this paper, a data-driven approach for laser failure mode detection and remaining useful life (RUL) prediction is developed using Long Short-Term Memory (LSTM) based methods. The proposed approach is demonstrated and validated using synthetic reliability data. The different laser degradation modes are detected with higher accuracy than with conventional techniques, and the RUL is accurately predicted. The presented framework provides a generalizable approach for laser monitoring and early failure prediction during operation by using current sensor measurements and operating conditions. The remainder of the paper is organized as follows. Section 2 describes the proposed approach and the subsections detail the different algorithms incorporated in the diagnostics and prognostics. The validation of the presented framework using simulations is shown in Section 3. Conclusions are drawn in Section 4.

2. **ML-based Laser Diagnostics and Prognostics Framework**

The proposed ML based framework for laser diagnostics and prognostics includes mainly three stages: data acquisition and preprocessing of the collected data, failure detection and RUL prediction once a fault is detected, as shown in the flow chart of Figure 1. In the first step, we acquire and pre-process the data. Then the pre-processed data is fed to the LSTM-based fault detection model to identify the laser health status as a) normal, b) suddenly degrading or c) gradually degrading. If the laser is in normal operation, it is infeasible to predict the RUL. If the laser is degrading, we use the LSTM-based RUL prediction model to estimate RUL. The following subsections explain in more detail the different steps.

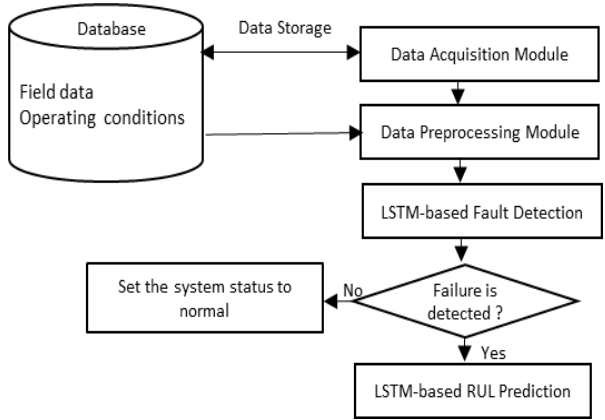
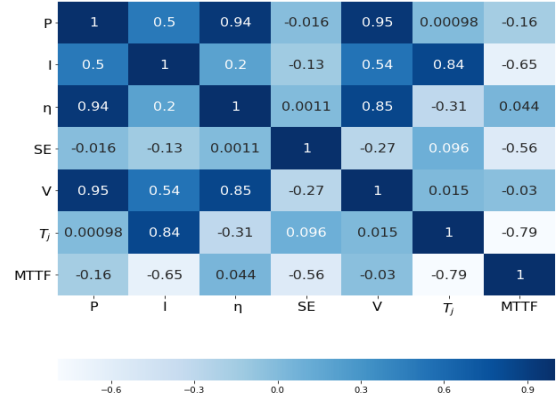

*Figure 1. Flow chart of the ML based laser diagnostics and prognostics framework*   *Figure 2. Reliability data correlation matrix*

## 2.1 Data Acquisition and Pre-Processing

To estimate the laser reliability, it is necessary to monitor different laser degradation parameters, namely junction temperature $T_j$, optical power $P$, forward current $I$, slope efficiency $SE$, conversion efficiency $\eta$ and voltage $V$. To reduce the number of parameters to be controlled, a correlation-based feature selection method is applied to synthetic reliability data, representing the mean time to failure MTTF (i.e. reliability metric) as a function of different laser degradation parameters. See [4] for more details on the process of synthetic reliability data generation. The correlation matrix shown in Figure 2 yielded that the two most important parameters influencing the laser reliability are the junction temperature and the current. Considering that those two parameters are highly correlated and that accurately measuring the p-n junction temperature is difficult, the current can be used as the primary laser performance indicator to monitor the degradation in real-time.

The current can be measured periodically using a current sense resistor (i.e. a current sensor). The measured values are stored in a database. Please note that the laser degradation modes are observed at different time scales: gradual degradation could extend to several hundreds of hours, and catastrophic degradation appears after many hours of normal operation, an accurate prediction requires the combination of the recent measurements representing the latest change and the historical sensor measurements jointly modelling the degradation tendency. As the sensor measurements are noisy, a Savitzky-Golay filter [5], performing data smoothing based on least squares polynomial approximation, is applied. The denoised sensor measurements are normalized to a window size of 100. The pre-processed measurement data is combined with the laser operating conditions, influencing the failure criteria, namely temperature, optical power and wavelength. Figure 3 shows the process of data acquisition and preprocessing.

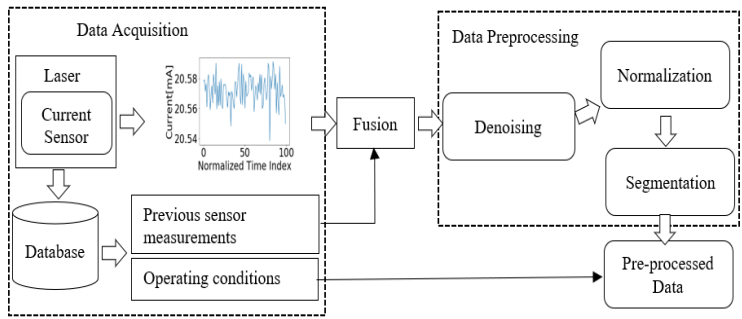
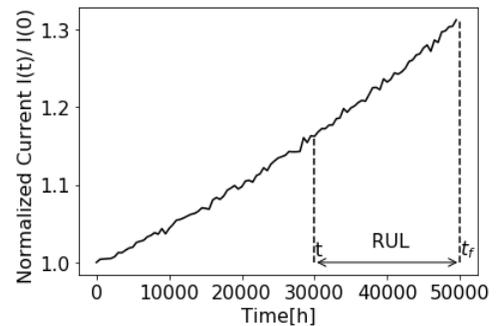

*Figure 3. Data acquisition and preprocessing process*   *Figure 5. Definition of RUL, $t$ is the current time, $t_f$ the time of failure*

## 3. LSTM-based Fault Detection

In our previous work [6], we proposed a data-driven fault detection model based on Long Short-Term Memory (LSTM) to detect the different laser failure modes, namely rapid, sudden and gradual degradation, modelling the different laser degradation patterns as well as normal laser operation with synthetic data. Given the current sensor data combined with laser operating conditions, namely current threshold, temperature, optical power and wavelength, the LSTM model detects the type of laser degradation. In this paper, the performance of the proposed model in terms of accuracy is improved and the architecture is optimized. Considering that the rapid degradation is observed only within the first

100 hours of operation, we excluded this type of degradation from the classes/targets that the model predicts. On the basis of the model performance analysis as a function of the operating conditions, we decided not to consider the current threshold. Therefore, the input of the model is the current measurement data combined with the laser parameters temperature, optical output power, and wavelength, and the output is the laser degradation type (normal, sudden degradation and gradual degradation). The model consists of one hidden layer with 50 neurons. The general structure of the proposed LSTM based fault detection model is depicted in Figure 4.

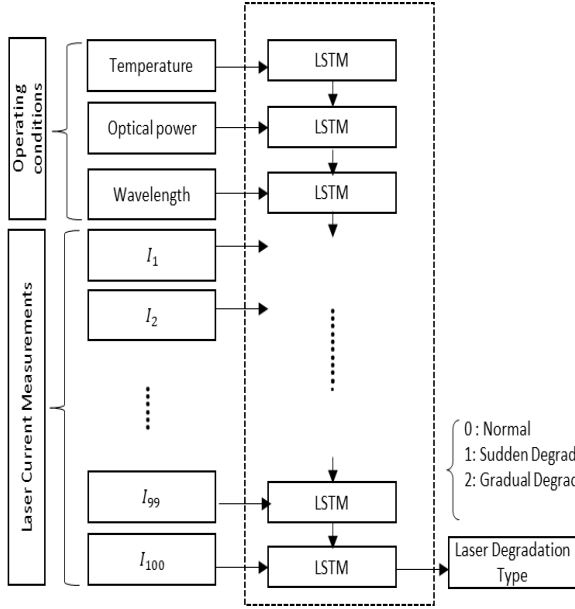

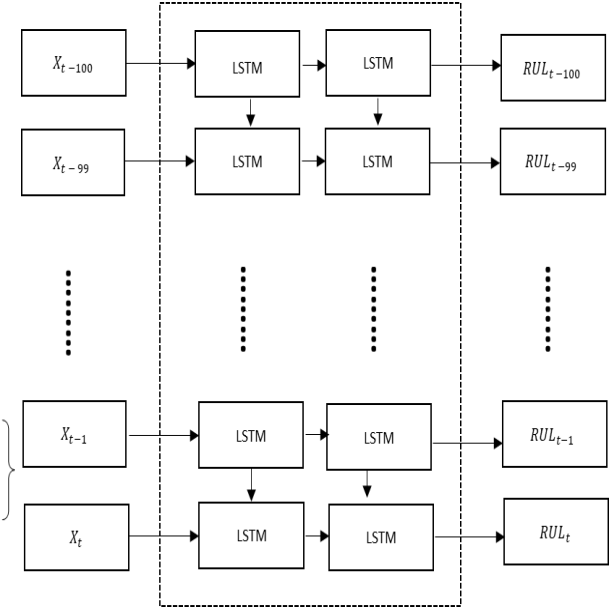

*Figure 4. LSTM-based fault detection architecture*     *Figure 6. LSTM-based RUL prediction architecture*

### 3.1 LSTM-based RUL Prediction

After the laser degradation mode detection, it is crucial to estimate the RUL of the device, defined as the length of time a device is likely to operate before being repaired or replaced, as shown in Figure 5. To accurately estimate the RUL, it is necessary to build a model able to capture the time series sequence information in the current sensor data. As LSTM is good at temporal modelling, it is well suited for this problem. The model is trained by a dataset including multiple run-to-failure data sequences. Each run-to-failure data sequence is a univariate time series in form of $X = [X_{t-N}, X_{t-N-1}, \ldots X_{t-1}, X_t]$ where each data sample is a vector containing the current sensor value at that sample time along with the operating conditions. For each data sample, the LSTM model predicts the RUL at that sample time. To specify the RUL label for each data sample, a piece-wise labelling approach assuming that the laser performance starts to degrade linearly at some point $\tau$ is used. The formula used to label the RUL in form of normalized life percentage is given by equation (1), where $t_f$ denotes the time to failure of the device.

$$RUL\ (\%) = \begin{cases} 1 & \forall\ t \leq \tau \\ \dfrac{t_f - t}{t_f - \tau} & \forall\ t > \tau \end{cases} \quad (1)$$

Given the different time scales of sudden and gradual degradation and hence the different sampling rates of the current sensor measuremnts, two LSTM-based RUL prediction models are developed for sudden and gradual degradation respectively. Each model has a stacked-LSTM architecture composed of two hidden layers. The general structure of each model is shown in Figure 6.

### 4. Validating the proposed framework

### 4.1 Synthetic Current Sensor Data Generation

To validate the proposed framework, we modeled the current sensor measurement data using equation (2), showing the variation of the operational current as a function of the time at constant power [7].

$$I(t) = I_0 + I_{nr}(t) + N(\mu, \sigma^2)\ where\ I_{nr}(t) = \beta \exp(kt)\ and\ k = P^n \exp(\mu_0 - \frac{E_A}{k_B T}) \quad (2)$$

The parameter *n* denotes the de-rating exponent, $E_A$ the activation energy, $\mu_0$ the scale parameter, $\beta$ the non-radiative current and *T* the temperature. $N(\mu, \sigma^2)$ represents the Gaussian noise of mean µ and variance $\sigma^2$. The input features, including optical power *P*, the threshold current $I_0$ and temperature *T*, are extracted from real laser datasheets specifications, whereas the underlying coefficients to create the different degradation patterns are generated using normal distributions.

A syntethic data for sudden and gradual degradation types as well as the normal laser behavior is generated. The said data is denoised and normalized to a window size of 100. The current samples combined with laser operating conditions are fed to the LSTM models for training and evaluation.

### 4.2 Results

The results show that the LSTM-based fault detection model detects the different laser failure modes with a high accuracy of up to 98.1%. The confusion matrix depicted in Figure 8, underlines that the class "sudden degradation" is in rare cases misclassified as normal. This can be explained by the fact that the sudden degradation trend is smiliar to normal behavior with suddenly increasing tendency within short time at the end of life of the device. Hence the abrupt degradation pattern is hidden by the large amount of normal behavior data within the same data sample. The results demonstrate as well that the LSTM-based RUL prediction model for sudden degradation predicts the RUL with an RMSE (root mean square error) up to 36 min. For the gradual degradation, the RUL prediction error is estimated to be up to 131 hours. The plots in Figure 9 demonstrate that the predicted RUL trends to be close to the true piecewise RUL, especially during the end-of life stage.

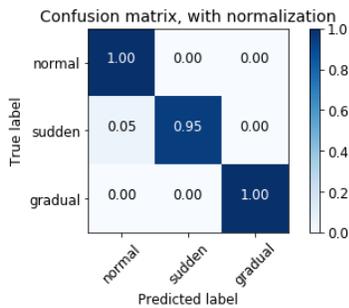
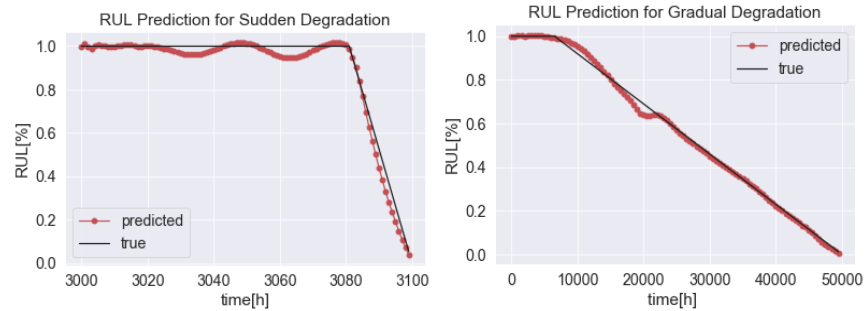

*Figure 8. Confusion matrix for fault detection*    *Figure 9. The true and the predicted RUL for randomly selected test samples*

### 5. Conclusions

In this paper, we proposed a data-driven perspective to enhance the laser relaibility using ML techniques. The proposed approach detects the laser failure modes and predicts the RUL with high accuracy. Future work will include the validation of the presented framework using in-field data and combine the two LSTM models for failure detection and RUL prediction into one single model using multi-task learning.